\documentclass[sigconf]{acmart}

\renewcommand\footnotetextcopyrightpermission[1]{} 

\usepackage{comment}
\usepackage{amsmath,amssymb,amsfonts}
\DeclareMathOperator*{\argmin}{argmin}
\DeclareMathOperator*{\argmax}{argmax}
\usepackage{algorithmic}
\usepackage{graphicx}
\usepackage{textcomp}
\usepackage{xcolor}
\usepackage{multirow}
\usepackage{float}
\usepackage{tikz}
\usepackage{circuitikz}
\usepackage{caption}
\usepackage{svg}
\usepackage{subfig}
\usepackage{hyperref}
\usepackage{balance}
\usepackage[linesnumbered, ruled, vlined]{algorithm2e}
\let\oldnl\nl
\newcommand{\nonl}{\renewcommand{\nl}{\let\nl\oldnl}}

\newcommand\encircle[1]{%
\tikz[baseline=(X.base)]
 \node (X) [draw, scale=0.75, shape=circle, inner sep=0, fill=black, text=white, minimum size=0em] {\strut #1};}

\AtBeginDocument{%
  }

\hypersetup{
    colorlinks=true,
    linkcolor=black,
    urlcolor=black,
    filecolor=black,
    citecolor=black
}

\begin{document}
\fancyhead{}
\title{CrossNAS: A Cross-Layer Neural Architecture Search Framework for PIM Systems}


\author{Md Hasibul Amin}
\affiliation{%
  \institution{University of South Carolina}
  \city{Columbia}
  \state{SC}
  \country{USA}}
\email{ma77@email.sc.edu}

\author{Mohammadreza Mohammadi}
\affiliation{%
  \institution{University of South Carolina}
  \city{Columbia}
  \state{SC}
  \country{USA}}
\email{mohammm@email.sc.edu}

\author{Jason D. Bakos}
\affiliation{%
  \institution{University of South Carolina}
  \city{Columbia}
  \state{SC}
  \country{USA}}
\email{jbakos@cse.sc.edu}

\author{Ramtin Zand}
\affiliation{%
  \institution{University of South Carolina}
  \city{Columbia}
  \state{SC}
  \country{USA}}
\email{ramtin@cse.sc.edu}


\begin{abstract}
In this paper, we propose the CrossNAS framework, an automated approach for exploring a vast, multidimensional search space that spans various design abstraction layers—circuits, architecture, and systems—to optimize the deployment of machine learning workloads on analog processing-in-memory (PIM) systems. CrossNAS leverages the single-path one-shot weight-sharing strategy combined with the evolutionary search for the first time in the context of PIM system mapping and optimization. CrossNAS sets a new benchmark for PIM neural architecture search (NAS), outperforming previous methods in both accuracy and energy efficiency while maintaining comparable or shorter search times.
\end{abstract}



\maketitle

\section{Introduction}

Processing-in-memory (PIM) architectures have emerged as promising alternatives to conventional von Neumann-based machine learning (ML) hardware \cite{IMCSurvey}. These architectures exploit features such as massive parallelism, analog computation, and the ability to perform computations directly where data is stored, leading to significant performance improvements \cite{PUMA,prime,ISAAC,TPU-IMAC}. The foundation of most PIM architectures is based on memristive crossbar arrays, which utilize resistive memory technologies such as resistive random-access memory (RRAM) and magnetoresistive random-access memory (MRAM) \cite{zand2018fundamentals}. These arrays enable matrix-vector multiplication (MVM) in the analog domain using basic circuit laws \cite{iCASISVLSI22,dpengine}.


\begin{figure}[h!] 
\centering
\includegraphics [width=0.98\linewidth]{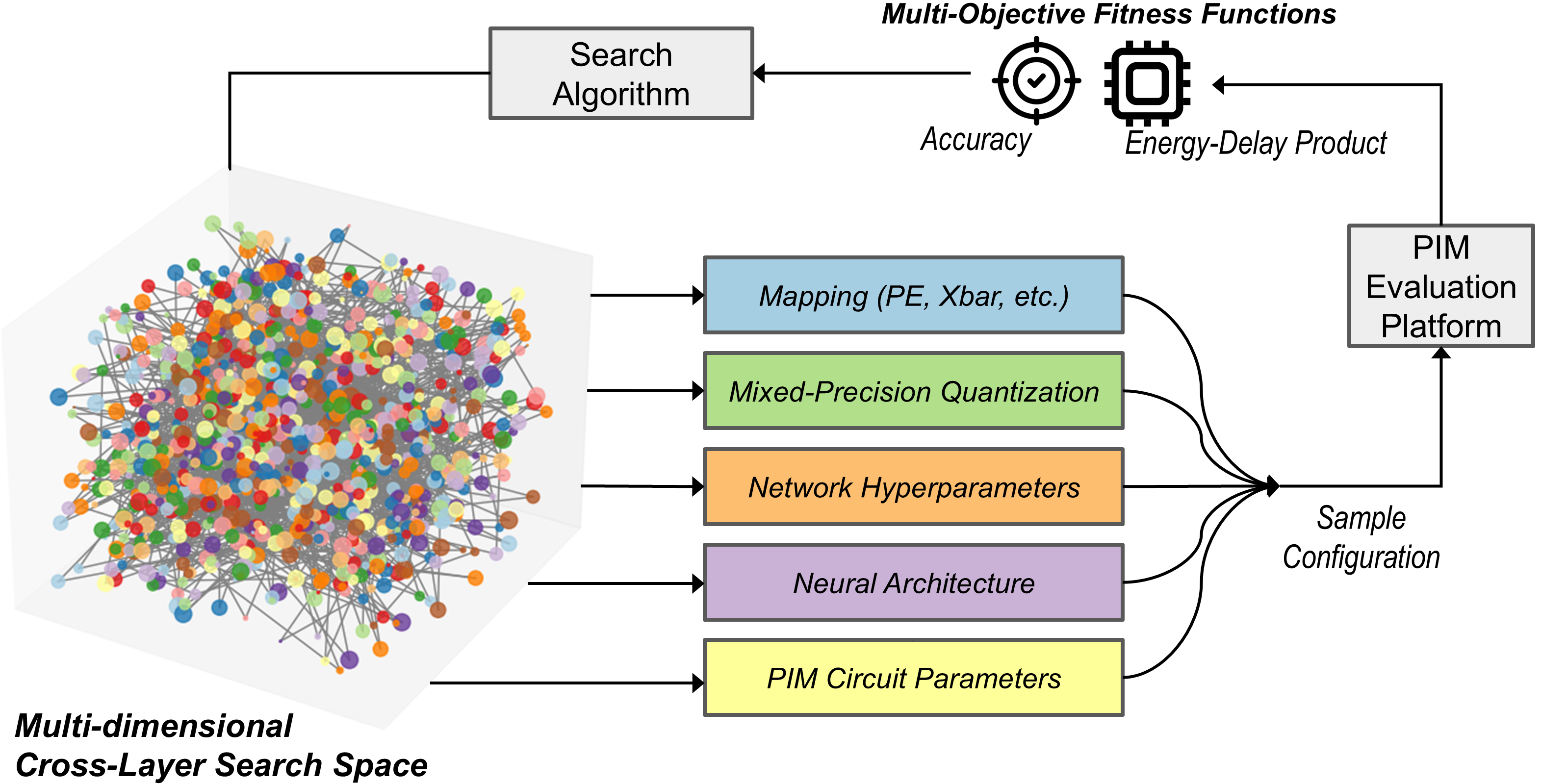}
\caption{The CrossNAS method explores a multi-dimensional search space across multiple design abstraction layers—circuit, architecture, and system levels—using a custom multi-objective search and optimization process.}
\label{fig:motivation}
\end{figure}

\begin{figure*}[]
    \centering
    \includegraphics[width=0.95\textwidth]{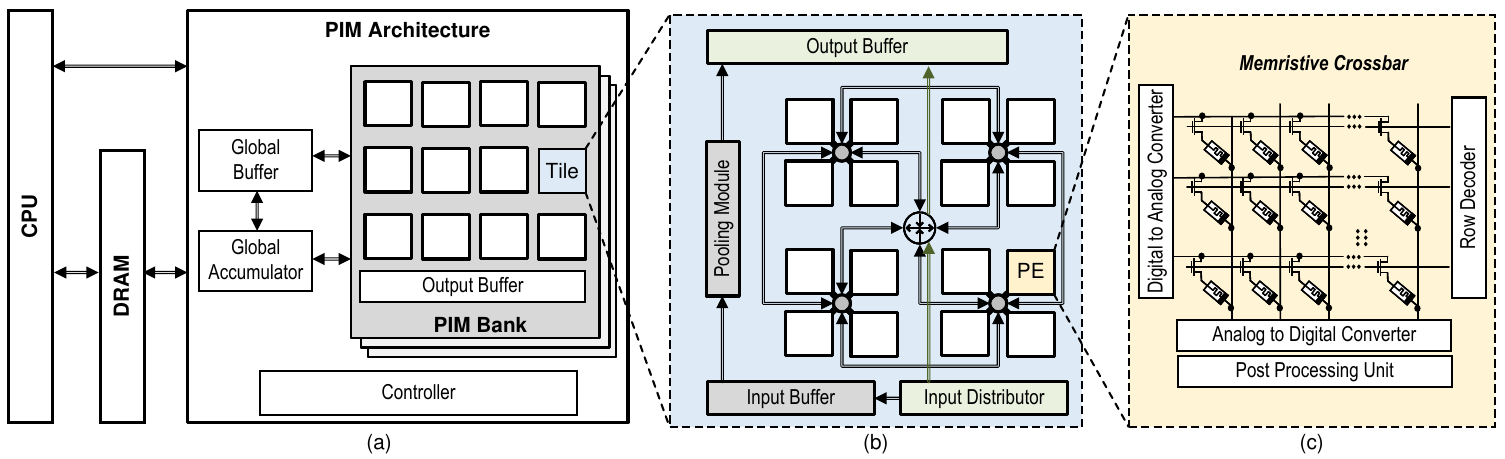}
    \vspace{-3mm}
    \caption{(a) The analog PIM architecture with multiple banks including several interconnected tiles. (b) The PIM tile consists of a network of processing elements (PEs). (c) The PEs include memristive crossbars and signal conversion units.}
    \label{fig:arch}
    \vspace{-3mm}
\end{figure*}

Despite the progress in PIM architectures, previous research indicates that deploying pre-trained ML models, designed and optimized for digital von Neumann architectures, does not consistently achieve comparable performance on analog PIM architectures \cite{xbarpartition}. This is caused by several factors, including the limited numerical precision of memristive devices \cite{mixedIMC}, and circuit imperfections such as interconnect parasitics \cite{parasiticsiCAS}, and device variations \cite{variation}. 



An alternative strategy for mapping and deploying ML models to PIM architecture includes the automated tuning of neural architecture parameters alongside PIM circuit and device-level hyperparameters to achieve specific design objectives \cite{10682726}. NACIM \cite{nacim} and UAE \cite{uae} provide such exploration frameworks based on reinforcement learning and LSTM controller, respectively, but they explore only VGG-like models, which limits their performance. NAS4RRAM \cite{nas4rram} also provides a similar framework for ResNet-like architectures. NACIM, UAE and NAS4RRAM need to train the sampled architectures from scratch to estimate the accuracy which increases the search time. NAX \cite{nax} trains an over-parameterized network for both architecture and hardware optimization but their architecture search space only contains filter sizes for each layer, lacking block types or channel numbers. Gibbon \cite{gibbon} uses a recurrent neural network (RNN)-based predictor for estimation of accuracy and hardware performance, but it only includes ResNet-like blocks in their architecture. AnalogNAS \cite{analogNAS} trains a surrogate model which predicts the behavior of searched architectures but they mostly focus on ResNet-like architectures and PIM configuration parameters are not included in the search space.

In this work, we present the CrossNAS framework, an automated approach for exploring a vast multidimensional search space across PIM design abstraction layers, including circuit, architecture, and system levels, as shown in Fig. \ref{fig:motivation}. We propose a weight-sharing-based neural architecture search (NAS) approach to explore the cross-layer search space, encompassing diverse neural building blocks, convolutional channel configurations, layer-specific quantization, and PIM circuit-level hyperparameters.

The main challenge of the weight-sharing approaches is to overcome the weight coupling among different architectures within the supernet, which raises concerns about the effectiveness of inherited weights for a specific architecture. Various weight-sharing approaches have been previously explored for neural architecture search \cite{wsnas-basic, fbnet, oneshotzoph, proxyless, spos, ssnas, spos-hw}. Many weight-sharing methods employ continuous relaxation to parameterize the search space, where architecture distribution parameters are jointly optimized during supernet training using gradient-based techniques \cite{wsnas-basic, fbnet}. After optimization, the best architecture is sampled from the distribution. However, the joint optimization method entangles architecture parameters with supernet weights, and the greedy nature of gradient-based methods introduces bias, potentially misleading the architecture search. The one-shot paradigm \cite{oneshotzoph, proxyless} alleviates this issue by decoupling the architecture search from supernet training. It defines the supernet and performs weight inheritance similarly, without architecture relaxation.
However, the issue of weight coupling among different architectures still persists. To overcome this issue, single-path \cite{spos, ssnas, spos-hw} NAS has been proposed, where all architectures are optimized simultaneously, through uniform sampling during supernet training.



The CrossNAS framework utilizes the single-path one-shot \cite{spos} weight-sharing strategy to find the optimum neural model and corresponding PIM parameters. The process involves training an overparameterized supernet based on neural search space, followed by using an evolutionary algorithm to select optimal subnets based on multi-objective fitness functions. 
The contributions of our work can be summarized as follows:


\begin{itemize}
    \item Creating a multi-dimensional cross-layer search space including different architectures, channel counts, layer-specific weight and activation bit widths, as well as PIM parameters such as crossbar size and the precision of analog-to-digital (ADC) and digital-to-analog (DAC) converters. We introduce new parameters to the PIM search space, allowing for the selection of different neural building blocks for each layer of the model.
    \item Adapting the single-path one-shot weight-sharing strategy along with the evolutionary search for the first time in the context of mapping and optimization of PIM systems.    
    \item Establishing a new benchmark for PIM neural architecture search that surpasses prior works in terms of accuracy and energy efficiency.
    
\end{itemize}

\section{PIM Architecture}

Figure \ref{fig:arch} illustrates the structure of the analog PIM architecture utilized in this work \cite{MNSIM2}. This architecture features multiple PIM banks, a global buffer, a global accumulator, and a PIM controller, as shown in Fig. \ref{fig:arch} (a). The PIM controller oversees data transfer between DRAM and PIM banks and receives status updates from the CPU. The global buffer and global accumulator support elementwise summation, which is crucial for implementing skip connections in ML models such as ResNet \cite{resnet}.

Each PIM bank is structured into arrays of tiles interconnected via a network-on-chip (NoC). Within each tile are processing elements (PEs), a pooling module, and input/output buffers,  as depicted in Fig. \ref{fig:arch} (b). The PEs include memristive crossbars and peripheral circuits such as ADCs, DACs, and a post-processing unit for handling nonlinear vector operations. The memristive crossbars execute MVM operations in the analog domain by leveraging basic circuit principles: Ohm's law is used for multiplication ($I=GV$), while Kirchhoff’s law enables accumulation via charge conservation \cite{xbarpartition,dpengine}. The weight kernels are expanded into a vector and loaded onto crossbar columns by adjusting the conductance ($G$) of memristive devices, while the input feature maps are applied as input voltages ($V$) to the crossbar. In this work, we use MNISIM 2.0 \cite{MNSIM2} to emulate PIM circuit behavior and map ML models onto PIM hardware, allowing us to measure model accuracy, as well as PIM energy consumption and latency. 

\section{The Proposed CrossNAS Methodology}

\begin{figure}[] 
\centering
\includegraphics [width=0.98\linewidth]{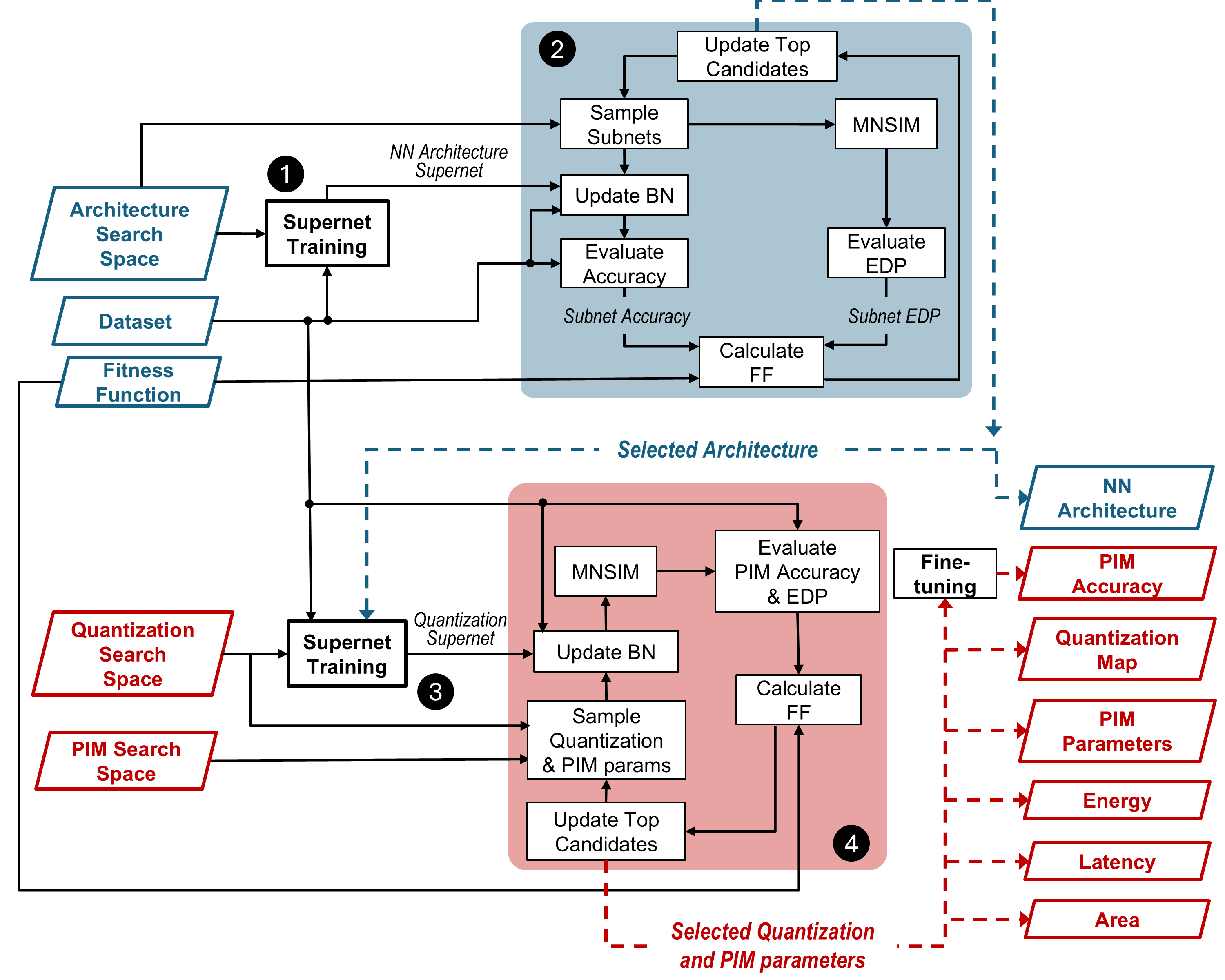}
\caption{Optimization flow for the CrossNAS framework.}
\vspace{-4mm}
\label{fig:flow}
\end{figure}

Figure \ref{fig:flow} provides a high-level description of the proposed CrossNAS framework which includes four main steps. \encircle{1} Train an overparameterized supernet comprising all possible architectures in the search space using a single-path one-shot weight-sharing strategy. \encircle{2} Use an evolutionary algorithm to search subnet architectures and evaluate them based on a multi-objective fitness function (FF) comprising accuracy and performance metrics. We use corresponding inherited weights from the supernet to approximate accuracy for the subnets, so no retraining is required. We adopt a behavior-level PIM simulator MNSIM 2.0 \cite{MNSIM2} for fast calculation of the hardware performance metrics. \encircle{3} The single-path one-shot weight-sharing strategy is utilized to train a mixed-precision quantization supernet based on the selected neural network (NN) architecture comprising all possible quantization scenarios in the quantization search space. \encircle{4} Using an evolutionary algorithm, we search the quantization space and PIM parameters to maximize FF. To calculate the FF, supernet weights are sent to MNSIM to measure the PIM-based NN accuracy and hardware performance metrics for the candidate quantization map and PIM parameters.

The single-path one-shot NAS method is described in section \ref{sec:spos-nas}. The supernet training and subnet search methods for NN architectures (steps \encircle{1} and \encircle{2})  are described in section \ref{sec:arch}. Finally, section \ref{sec:quant} describes the mixed-precision quantization supernet training along with the evolutionary search of quantization space and PIM hyperparameters (steps \encircle{3} and \encircle{4}).

\subsection{Single-path One-shot NAS}
\label{sec:spos-nas}

According to the single-path one-shot NAS approach \cite{spos}, the supernet weight training and the subnet architecture search are performed in two separate sequential steps. The supernet is an overparameterized network which contains the trained weights for all possible architectures in the search space. The subnet is a subset of the supernet which is evaluated using its inherited weights from the supernet. For the supernet training, our goal is to minimize the training loss for all possible architecture choices in the search space as shown in equation (\ref{eq:sup}),

\begin{equation}
    W_{SUP}=\argmin_{W}\mathcal{L}_{train}(\mathcal{N}(A,W))
    \label{eq:sup}
\end{equation}

\noindent where $A$ is the architecture search space, $\mathcal{N}(A,W)$ represents the set of all subnet architectures, $\mathcal{L}_{train}$ is the loss function. To ensure that the accuracy of a sampled architecture $a$ on the test dataset using inherited weights $W_{SUP}(a)$ is highly predictive of the accuracy of fully-trained architecture $a$, we simultaneously optimize the weights for all architectures in the search space through uniform sampling. So, equation (\ref{eq:sup}) can be rewritten as,

\begin{equation}
    W_{SUP}=\argmin_{W}\mathbb{E}_{a \sim \Gamma(A)}[\mathcal{L}_{train}(\mathcal{N}(a,W_{SUP}(a)))]
    \label{eq:supm}
\end{equation}

\noindent where $\Gamma(A)$ is a prior distribution of $a\epsilon A$. Thus, in each optimization step, we randomly sample one architecture $a$ and only its corresponding weight $W_{SUP}(a)$ is activated and trained. Therefore, By the end of training, the supernet acts as an approximate model, representing the weights of fully-trained subnet architectures.

Once the supernet is trained, we sample subnets from the trained supernet architecture and evaluate them based on the desired FF. During the subnet search, each of the sampled architecture inherits weights from the supernet $W_{SUP}$ as $W_{SUP}(a)$. The best architecture is selected as,

\begin{equation}
    a^{*}=\argmax_{a\epsilon A}FF_{test}\mathcal{N}(a,W_{SUP}(a))
    \label{eq:sub}
\end{equation}


Since each fitness function ($FF_{test}$) utilized in this work includes accuracy as a key metric, we assess accuracy through inference using weights inherited from the supernet. With access to these pre-trained supernet weights, our search process becomes highly efficient, eliminating the need for retraining. To search for optimized subnets based on specific FFs, we apply an evolutionary algorithm as described in the following.


\begin{figure}[] 
\centering
\includegraphics [width=0.98\linewidth]{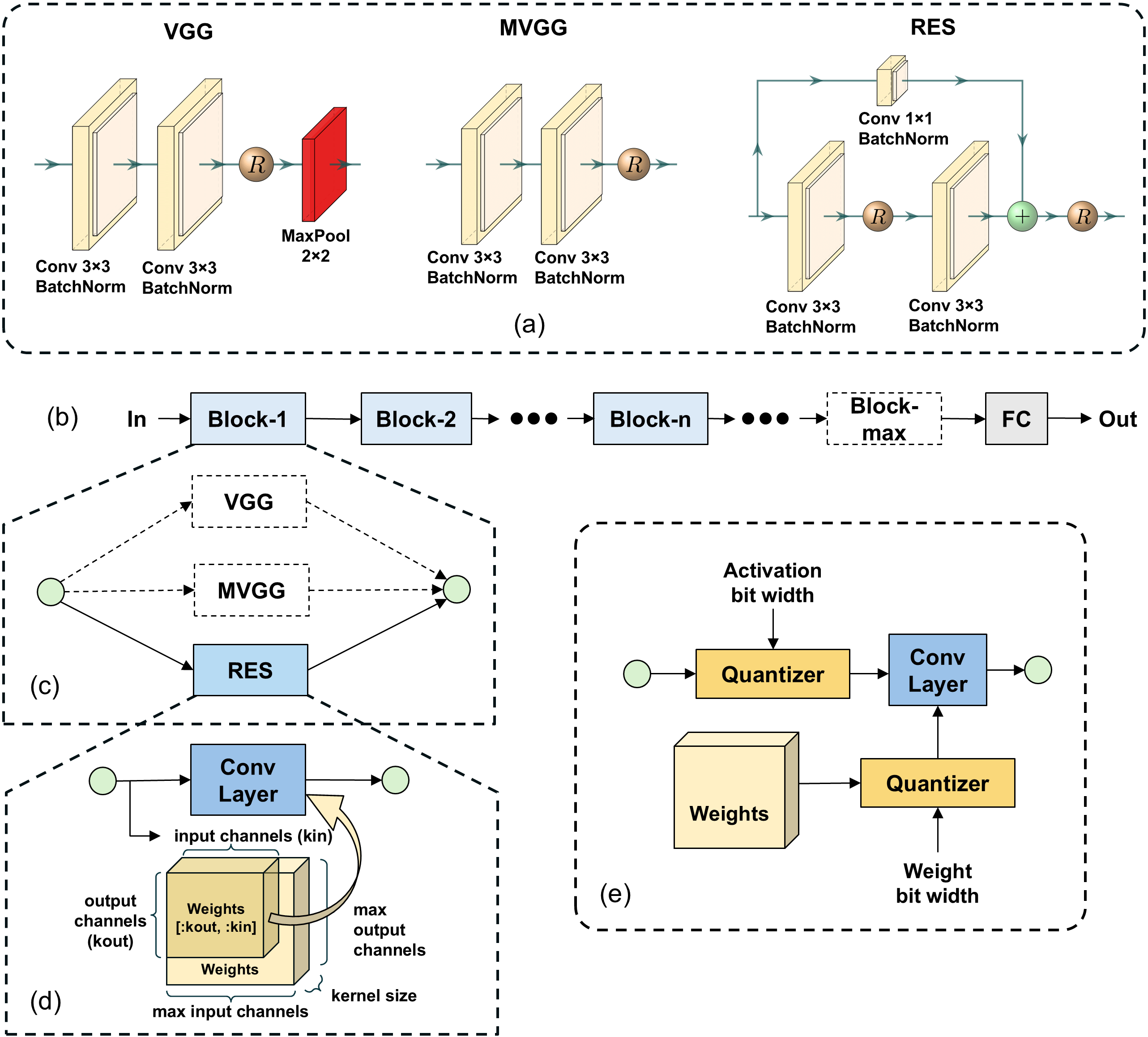}
\caption{Hierarchical sampling of subnet from supernet. (a) building blocks, (b) NN architecture buildup by depth sampling, (c) block sampling, (d) output channel sampling via slicing of the weight matrix, and (e) layerwise weight and activation quantization}
\vspace{-4mm}
\label{fig:super}
\end{figure}

\subsection{Neural Network (NN) Architecture Search}
\label{sec:arch}


\begin{table}[t]
\centering
\caption{Network Configuration Settings}
\vspace{-3mm}
\begin{tabular}{cc}
\hline
\multicolumn{2}{c}{\textbf{Architecture Search Space}} \\
\hline
No. of blocks in the network ($n$) & 1-8            \\
The block type ($Block$)              & VGG, MVGG, RES     \\
No. of out. channels in each block ($K$) & 32, 64, 128        \\ \hline
\multicolumn{2}{c}{\textbf{Quantization Search Space}} \\
\hline
Weight bit width ($WB$) & 5, 7, 9      \\
Activation bit width ($AB$) & 5, 7, 9  \\   \hline
\multicolumn{2}{c}{\textbf{PIM Search Space}} \\
\hline
Crossbar size ($Xbar$) & 32, 64, 128, 256     \\
ADC resolution ($ADC$)  & 4, 6, 8, 10       \\
DAC resolution ($DAC$)  & 1, 2        \\
\hline
\end{tabular}
\label{tab:space}
\vspace{-3mm}
\end{table}

As shown in Figure \ref{fig:super}, we build an overparameterized supernet containing the weights for all possible architecture choices in the search space. As building blocks for our network configuration, we design three different convolutional choice blocks based on the well-known VGG \cite{vgg16} and ResNet \cite{resnet} architectures, as shown in Fig. \ref{fig:super} (a). The three building blocks are: (1) the VGG block with two $3\times3$ convolution layers with a stride of $1$ followed by a $2\times2$ max pooling layer, (2) the modified VGG block (MVGG) which is similar to VGG without the sub-sampling layer, and (3) the RES block with two $3\times3$ convolution layers and a residual connection with a $1\times1$ convolution block. The ReLU activation function is used as the nonlinear unit in all of the building blocks. The full search space is shown in Table \ref{tab:space}.


As shown in Fig. \ref{fig:super} (b), the supernet is configured with the maximum depth ($D_{max}$) and includes the highest possible number of sequential convolutional (conv) blocks within the search space, followed by fully connected layers as classification heads. Each conv block in the supernet includes all three types of VGG, MVGG, and RES blocks connected in parallel paths (Fig. \ref{fig:super} (c)). The size of the conv layers weight matrices in the supernet are set according to maximum input ($k_{in}$) and output ($k_{out}$) channel sizes possible (Fig. \ref{fig:super} (d)). At each training step, the algorithm randomly selects a depth of $n$ blocks, one of the VGG, MVGG, or RES paths in each block, and output channel count ($k_{out}$) for each block. Based on this selection, the first $n$ conv blocks and their corresponding paths are activated, and the remaining $D_{max}-n$ blocks are skipped. The weight matrices of active conv layers and paths are then sliced according to the selected input and output channels, updating only the relevant kernel portions in that training step.

During the search phase, we employ an evolutionary algorithm to identify top subnet candidates based on a multi-objective FF. We introduce a PIM-oriented FF, defined as a weighted sum of accuracy and the normalized energy-delay product (EDP), as shown in equation (\ref{eq:FF}).
\begin{equation}
    FF=w_{acc}\times accuracy - (1-w_{acc})\times EDP_{norm}
    \label{eq:FF}
\end{equation}

\noindent where $w_{acc}$ can be tuned to any number between 0 and 1 to adjust the balance between accuracy and EDP.

We utilize the MNSIM 2.0 \cite{MNSIM2} simulator to determine the energy and latency of the selected subnet model's PIM implementation. The $w_{acc}$ is set based on user preference. Since the accuracy of each subnet—obtained using the inherited supernet weights—offers a reliable estimate for comparing subnet architectures, we use this approximate accuracy rather than performing the more time-intensive PIM-specific accuracy measurement in this phase. Because the supernet’s batch normalization (BN) statistics do not apply to candidate subnets, we retrain only the BN layers on the training set before evaluating subnet accuracy via inference. This BN retraining takes just a few seconds, adding minimal computational or time cost to the process.


We employ Algorithm \ref{algo:evol_sub} to identify the best-performing candidate subnet architecture from the search space. First, we initialize a candidate set $P_i$ with $P$ random subnets, which will represent the new population in each search cycle. Additionally, we maintain a set $topK$ to store the top $k$ subnet candidates with the highest fitness. After calculating the FF for candidates in $P_i$, we update $topK$ in descending order based on fitness scores. Based on the $topK$ candidates and a predefined mutation probability $prob$, we generate $m$ crossover and $n$ mutation candidates, with the newly generated candidates forming the updated $P_i$. This process is repeated for a fixed number of cycles ($cyc$). Ultimately, the candidate with the highest fitness value is selected as the optimal architecture.

\begin{algorithm}[t]
\DontPrintSemicolon
\SetAlgoLined
\KwIn{supernet weights ($W_{SUP}$), training data ($train$), test data ($test$), population ($P$), maximum cycle ($cyc$), top k candidates ($topK$), mutation value ($n$), crossover value ($m$), mutation probability ($prob$)}
\KwOut{The NN architecture with the best value of corresponding fitness function}
\BlankLine

 \textbf{Initialize:} $P=50$, $P_0$=random P candidates, $topK=\phi$, $n=P/2$, $m=P/2$, prob=0.1\;

 \For{$i=1$ \KwTo $cyc$}{
  $FF_{i-1}$ $\Leftarrow$ Calculate\_FF($W_{SUP}$, $train$, $test$, $P_{i-1}$)\;
  topK $\Leftarrow$ Update\_topK(topK, $P_{i-1}$, $FF_{i-1}$)\;
  $P_{crossover}$ $\Leftarrow$ Crossover(topK, n)\;
  $P_{mutation}$ $\Leftarrow$ Mutation(topK, m, prob)\;
  $P_i$ $\Leftarrow$ $P_{crossover} \cup P_{mutation}$\;
  
}
Return the NN architecture with the highest value in topK
 \caption{Evolutionary Subnet Search}
 \label{algo:evol_sub}
\end{algorithm}

\subsection{Quantization and PIM Configuration Search}
\label{sec:quant}


We apply a single-path one-shot weight-sharing method to train a mixed-precision quantization supernet, followed by a search for optimal mixed-precision quantization and PIM configurations using an evolutionary algorithm. This search is performed on the optimal model architecture identified in the preceding NN architecture search step, which defines the supernet architecture for this stage. During training, weight and activation bit widths are randomly sampled from the search space at each step, and these sampled bit precisions are dynamically applied to the model, as shown in Fig. \ref{fig:super} (e). As a result, the supernet model approximates the fully trained mixed-precision quantized weights, allowing for efficient representation of all possible mixed-precision subnet configurations. 

We use a non-uniform quantization scheme \cite{PIM-quant} to quantize the weights and activations, as shown in the below equation.


\begin{equation}
    x_q=\text{\tt clip}\left(round\left(\theta \times \frac{x_{in}}{\alpha}\right),-\theta,\theta \right)\times\frac{\alpha}{\theta}
    \label{eq:quant}
\end{equation}



\noindent where $\alpha$ is the scaling parameter and $[-\theta,\theta]$ is the integer range. $\theta$ is calculated as $2^{q-1}-1$, where q is the quantization bit width. For weights, $\alpha$ is set to the maximum absolute value in the weight tensor. The activations have a broader range, which may include individual extreme values. So, for a fair reflection of the overall data distribution, we use $|\mu|+3|\sigma|$ as the scaling parameter for activations, where $\mu$ and $\sigma$ are the mean and standard deviation of the activation data, respectively. The scaling parameter $\alpha$ can be different in different mini-batches, which may cause jitter and non-convergence in training \cite{PIM-quant}. Therefore, each time a new $\alpha$ is generated, it is combined with the $\alpha$ from the previous mini-batch using a weighted sum, as shown in the equation (\ref{eq:alpha}).
\begin{equation}
    \alpha=m\alpha+(1-m)\times \left(|\mu\left(x_{in}\right)|+3|\sigma\left(x_{in}\right)|\right)
    \label{eq:alpha}
\end{equation}



To train a quantized model effectively, we begin by training a floating-point model and then fine-tune its weights to develop a mixed-precision quantization supernet. A fitness function and evolutionary algorithm, similar to those used in the NN architecture search step, are applied to explore various mixed-precision quantization mappings and PIM configurations. 
We update the batch normalization (BN) weights for the quantization subnet candidate before evaluating PIM-based NN accuracy in MNSIM. A lower mutation probability is assigned to the quantization map and a higher mutation probability to the PIM configuration parameters so that the top quantization candidates are cross-checked with different PIM configurations. Once the best subnet candidate is identified, we fine-tune the model for one last time to achieve optimal PIM accuracy.




\section{Results and Discussions}

\subsection{Experimental Setup}

The proposed CrossNAS framework is compatible with any PIM simulator that provides accuracy, energy, and delay metrics. For our experiments, we use MNSIM 2.0 \cite{MNSIM2} as the baseline simulator. The PIM accelerator is configured with a $64 \times 64$ tile arrangement, with each tile containing $2 \times 2$ PE arrays, and uses 1-bit memristor devices as the basic crossbar elements. We evaluate the performance of CrossNAS against previous exploration methods, including NACIM \cite{nacim}, UAE \cite{uae}, NAS4RRAM \cite{nas4rram}, and Gibbon \cite{gibbon}, on the CIFAR-10 and CIFAR-100 \cite{cifar} datasets. CIFAR-10 is selected because it serves as a common benchmark used in all previous related studies, allowing consistent comparisons. All experiments are conducted on a single NVIDIA{\textregistered} GeForce{\textregistered} RTX{\texttrademark} 2080 Ti GPU paired with an Intel{\textregistered} Core{\texttrademark} i9-9820X CPU at 3.30GHz.


\subsection{Training and Search Details}

During the neural network architecture search phase, we train the supernet using stochastic gradient descent (SGD) optimizer \cite{sgd} for 1000 epochs, starting with a learning rate of 0.1 and a batch size of 128. A learning rate scheduler is applied, reducing the learning rate by a factor of 5 every 250 epochs. The evolutionary algorithm runs for 10 cycles, evaluating 50 new candidates per cycle, consisting of 25 mutation candidates and 25 crossover candidates.


In the mixed-precision quantization and PIM configuration search phase, training begins with a floating-point (FP) model to achieve high accuracy. An SGD optimizer is used with an initial learning rate of 0.1, training for 200 epochs, with a step learning rate scheduler reducing the rate by a factor of 5 at epochs 60, 120, and 160. The FP weights are then used to retrain the model with variable mixed-precision quantization, employing the Adam optimizer \cite{adam} for quantization-aware training (QAT). Due to the sensitivity of QAT to learning rate, a low initial learning rate of 0.0008 is set, which is further divided by 5 every 40 epochs using a step learning rate scheduler.


\subsection{Performance Comparison Results on CIFAR-10 Dataset}

\begin{table}[]
\centering
\caption{Performance comparison against different PIM-oriented NAS methods on CIFAR-10 dataset.}
\vspace{-3mm}
\resizebox{\columnwidth}{!}{%
\begin{tabular}{ccccc}
\hline
\multirow{2}{*}{Method} & PIM & Latency & EDP & Search   \\
       & Accuracy & ($ms$) & ($mJ\times ms$) & time (h)  \\  \hline
NACIM \cite{nacim} & 73.9\% & - & 1.55 & 59  \\
UAE \cite{uae} & 83\% & - & - & 154  \\
NAS4RRAM \cite{nas4rram} & 84.4\% & - & - & 289 \\
Gibbon \cite{gibbon} (acc opt.) & 89.2\% & 3.44 & 1.67 & 6  \\
Gibbon \cite{gibbon} (edp opt.) & 83.4\% & 1.99 & 0.26 & 6  \\ \hline
CrossNAS ($w_{acc}$=0.99) & 91.27\% & 1.35 & 0.28 & 6\\
CrossNAS ($w_{acc}$=0.8) & 88.09\% & 0.577 & 0.073 & 5\\
\hline
\end{tabular}}
\label{tab:compare}
\end{table}

\begin{table}[]
\centering
\caption{Performance comparison against various well-known deep learning models on CIFAR-10 Dataset.}
\vspace{-3mm}
\resizebox{\columnwidth}{!}{%
\begin{tabular}{cccccc}
\hline
\multirow{2}{*}{Method} & PIM & Energy & Latency & EDP & Area   \\
       & Accuracy & (mJ) & (ms) & ($mJ\times ms$) & ($mm^2$)   \\  \hline
AlexNet \cite{alexnet}     & 81.7\% & 0.38 & 0.99 & 0.38 & 103.99      \\
VGG16 \cite{vgg16}       & 88.8\% & 2.68 & 6.43 & 17.22 & 499.57     \\
ResNet18 \cite{resnet}    & 89.7\% & 1.33 & 3.58 & 4.75 & 466.94\\ \hline
CrossNAS ($w_{acc}$=0.99) & 91.27\% & 0.21 & 1.35 & 0.28 & 306.64\\
CrossNAS ($w_{acc}$=0.8) & 88.09\% & 0.127 & 0.577 & 0.073 & 106.3\\
\hline
\end{tabular}}
\label{tab:c10}
\end{table}

\begin{figure}
    \centering
    \includegraphics[width=2.5in]{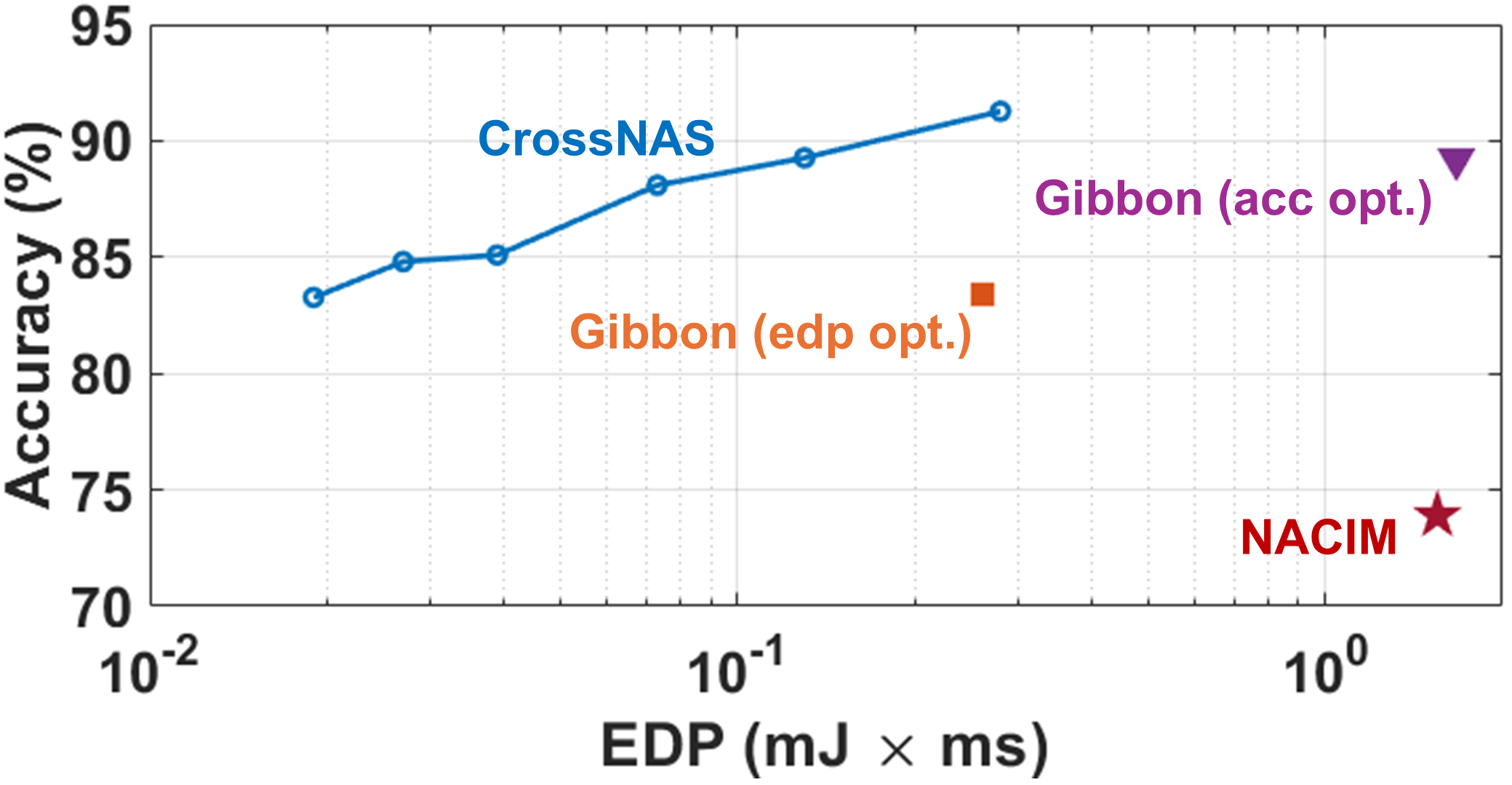}
    \caption{CrossNAS search results on CIFAR-10 dataset compared to previous PIM-oriented NAS methods.}
    \label{fig:sweep}
    \vspace{-3mm}
\end{figure}

Table \ref{tab:compare} presents a comparison of our results with previous PIM-oriented NAS frameworks on the CIFAR-10 dataset.  

To obtain a model optimized for high accuracy, we set the $w_{acc}$ to 0.99 in the FF. The model selected by CrossNAS at the end of optimization, shown in Fig. \ref{fig:models} (a), achieves an accuracy improvement of 2.07\% to 17.37\% over previous frameworks. The search process takes approximately six hours, comparable to Gibbon and up to 48.6$\times$ faster than other NAS methods. This model also achieves a $6\times$ reduction in energy-delay-product (EDP) compared to the model selected by Gibbon under accuracy optimization.

Next, we optimize the model for EDP by setting $w_{acc}$ to 0.8 in the FF, leading to the model shown in Fig. \ref{fig:models} (b). This EDP-focused model achieves a $3.6\times - 21.2\times$ reduction in EDP compared to previous methods, with only five hours of search time, and provides a 4.69\% accuracy improvement over the EDP-optimized model from Gibbon. Figure \ref{fig:sweep} shows the accuracy-EDP trade-off obtained for various $w_{acc}$ values in our FF. Overall, CrossNAS establishes a new benchmark for multi-objective optimization of PIM-based architectures, surpassing previous methods.



We also compare the models optimized by CrossNAS with other well-known deep learning models, such as AlexNet \cite{alexnet}, VGG16 \cite{vgg16} and ResNet18 \cite{resnet}, as presented in Table \ref{tab:c10}. The accuracy-optimized model achieves an accuracy improvement of 1.57\%-9.57\% compared to these models, along with an EDP reduction of $1.35\times - 61.5\times$. The EDP-focused model delivers an EDP reduction of $5.2\times - 235.9\times$, with a slight accuracy loss compared to VGG16 and ResNet18.


\subsection{Performance Comparison Results on CIFAR-100 Dataset}

\begin{table}[]
\centering
\caption{Performance comparison against various deep learning models on CIFAR-100 Dataset.}
\vspace{-2mm}
\resizebox{\columnwidth}{!}{%
\begin{tabular}{cccccc}
\hline
\multirow{2}{*}{Method} & PIM & Energy & Latency & EDP & Area   \\
       & Accuracy & (mJ) & (ms) & ($mJ\times ms$) & ($mm^2$)   \\  \hline
AlexNet \cite{alexnet}     & 57.1\% & 0.38 & 1.00 & 0.38 & 103.99      \\
VGG16 \cite{vgg16}       & 67.2\% & 2.68 & 6.44 & 17.25 & 499.57     \\
ResNet18 \cite{resnet}    & 72.3\% & 1.33 & 3.59 & 4.76 & 466.94\\ \hline
CrossNAS ($w_{acc}$=0.99) & 69.55\% & 0.81 & 3.62 & 2.93 & 282.12\\
CrossNAS ($w_{acc}$=0.8) & 60.09\% & 0.193 & 1.096 & 0.211 & 343.436\\
\hline
\end{tabular}}
\label{tab:c100}
\vspace{-4mm}
\end{table}

Here, we analyze the performance of CrossNAS in optimizing models for the CIFAR-100 dataset and compare the results with those of well-known deep learning models, as listed in Table \ref{tab:c100}. Similar to CIFAR-10, we set the $w_{acc}$ to 0.99 to obtain an accuracy-focused model, leading to the model shown in Fig. \ref{fig:models} (c). The accuracy-optimized model achieves a 12.45\% accuracy improvement over AlexNet, and a 2.35\% improvement over VGG16, but experiences a 2.75\% accuracy drop compared to ResNet18. The accuracy drop occurs because ResNet18 has channel numbers as high as 512, whereas our designed search space limits the maximum channel number to 128 to optimize EDP. Therefore, the accuracy-focused model achieves a $1.62 \times$ reduction in EDP compared to ResNet18. Additionally, by setting $w_{acc}$ to 0.8, we obtain an EDP-focused model, shown in Fig. \ref{fig:models} (d), which results in a $1.8\times - 22.6\times$ reduction in EDP compared to other models, along with a 3\% accuracy improvement over AlexNet. Note that a stride of 2 is used in the first $3 \times 3$ and $1 \times 1$ conv layers of the RES block when optimizing the models for CIFAR-100 by CrossNAS.




\subsection{Analysis of Selected Models}

\begin{figure}
    \centering
    \includegraphics[width=3.2in]{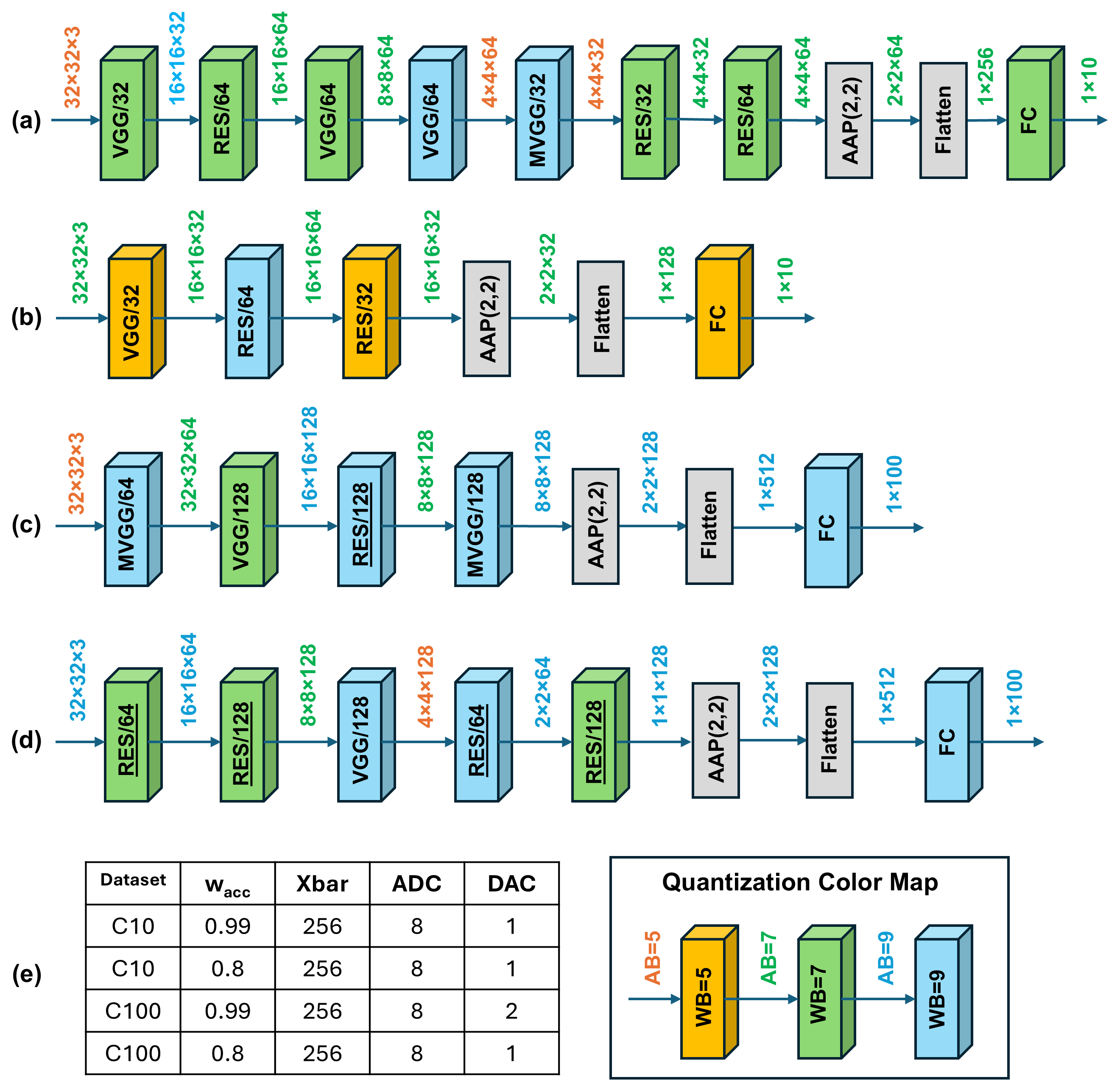}
    \vspace{-3mm}
    \caption{Block types along with channel numbers ($Block/K$) selected by CrossNAS for (a) CIFAR-10 dataset with $w_{acc}=0.99$ and (b) $w_{acc}=0.8$; (c) CIFAR-100 dataset with $w_{acc}=0.99$ and (d) $w_{acc}=0.8$; (e) corresponding PIM circuit configuration and quantization color map. AAP denotes adaptive average pooling, WB and AB denote weight and activation bit width, respectively. Underlined blocks indicate $stride=2$.}
    \label{fig:models}
    \vspace{-3mm}
\end{figure}

By analyzing various models optimized with different $w_{acc}$ values in the fitness function, we gain important insights that aid in designing PIM-specific model architectures tailored to particular design objectives.


Based on our observations from the optimized models selected by CrossNAS, shown in Fig. \ref{fig:models}, the EDP-focused models for the CIFAR-10 dataset typically select the VGG block as the first block, as it reduces the feature map size through max pooling. This reduction in feature map size lowers the hardware cost in subsequent layers, thereby reducing the overall EDP. For the CIFAR-100 dataset, the EDP-focused models tend to choose either the RES ($stride=2$) or VGG block as the first block, as both reduce the feature map size through higher stride or pooling layers.



From the mixed-precision quantization map, highlighted by various colors in Fig. \ref{fig:models}, we observe that the EDP-focused models selected by CrossNAS tend to use higher weight bit widths in the middle convolutional layers, while the head and tail layers have lower weight bit precision. The activation bit width is generally lower in the middle layers, whereas the head and tail layers exhibit higher activation precision. For shallower models, e.g., Fig. \ref{fig:models} (b), there is a tendency to apply the same activation bit width across the entire architecture.


From the PIM circuit configuration selected by CrossNAS, shown in Fig. \ref{fig:models} (e), we observe that the EDP-focused models prefer a crossbar size of $256 \times 256$. This configuration, combined with an 8-bit ADC, offers the best balance between accuracy and EDP.
Most of the EDP-focused models opt for a 2-bit DAC, resulting in minimal accuracy loss, while improving both latency and energy efficiency due to the reduced number of digital-to-analog conversions.


\section{Conclusion}

In this paper, we present CrossNAS, an efficient weight-sharing-based NAS framework for PIM systems. We construct a multi-dimensional cross-layer search space that includes diverse architectures, layer-specific weight and activation bit widths, and PIM parameters such as crossbar size and ADC/DAC precisions. Our search leverages a multi-objective fitness function to explore various neural network architectures and PIM hardware configurations. CrossNAS outperforms previous PIM-oriented NAS methods and most of the well-known deep learning models, identifying superior models in terms of both accuracy and EDP.


\begin{acks}
This work is supported in part by the National Science Foundation (NSF) under grant number 2409697.
\end{acks}

\balance
\bibliographystyle{ACM-Reference-Format}
\bibliography{Reference}


\begin{thebibliography}{34}


\ifx \showCODEN    \undefined \def \showCODEN     #1{\unskip}     \fi
\ifx \showISBNx    \undefined \def \showISBNx     #1{\unskip}     \fi
\ifx \showISBNxiii \undefined \def \showISBNxiii  #1{\unskip}     \fi
\ifx \showISSN     \undefined \def \showISSN      #1{\unskip}     \fi
\ifx \showLCCN     \undefined \def \showLCCN      #1{\unskip}     \fi
\ifx \shownote     \undefined \def \shownote      #1{#1}          \fi
\ifx \showarticletitle \undefined \def \showarticletitle #1{#1}   \fi
\ifx \showURL      \undefined \def \showURL       {\relax}        \fi
\providecommand\bibfield[2]{#2}
\providecommand\bibinfo[2]{#2}
\providecommand\natexlab[1]{#1}
\providecommand\showeprint[2][]{arXiv:#2}

\bibitem[Amin et~al\mbox{.}(2022a)]%
        {parasiticsiCAS}
\bibfield{author}{\bibinfo{person}{Md~Hasibul Amin}, \bibinfo{person}{Mohammed Elbtity}, {and} \bibinfo{person}{Ramtin Zand}.} \bibinfo{year}{2022}\natexlab{a}.
\newblock \showarticletitle{Interconnect Parasitics and Partitioning in Fully-Analog In-Memory Computing Architectures}. In \bibinfo{booktitle}{\emph{2022 IEEE International Symposium on Circuits and Systems (ISCAS)}}. \bibinfo{pages}{389--393}.
\newblock
\href{https://doi.org/10.1109/ISCAS48785.2022.9937884}{doi:\nolinkurl{10.1109/ISCAS48785.2022.9937884}}


\bibitem[Amin et~al\mbox{.}(2022b)]%
        {xbarpartition}
\bibfield{author}{\bibinfo{person}{Md~Hasibul Amin}, \bibinfo{person}{Mohammed~E. Elbtity}, {and} \bibinfo{person}{Ramtin Zand}.} \bibinfo{year}{2022}\natexlab{b}.
\newblock \showarticletitle{Xbar-Partitioning: A Practical Way for Parasitics and Noise Tolerance in Analog IMC Circuits}.
\newblock \bibinfo{journal}{\emph{IEEE Journal on Emerging and Selected Topics in Circuits and Systems}} \bibinfo{volume}{12}, \bibinfo{number}{4} (\bibinfo{year}{2022}), \bibinfo{pages}{867--877}.
\newblock
\href{https://doi.org/10.1109/JETCAS.2022.3222966}{doi:\nolinkurl{10.1109/JETCAS.2022.3222966}}


\bibitem[Amin et~al\mbox{.}(2024)]%
        {10682726}
\bibfield{author}{\bibinfo{person}{Md~Hasibul Amin}, \bibinfo{person}{Mohammadreza Mohammadi}, {and} \bibinfo{person}{Ramtin Zand}.} \bibinfo{year}{2024}\natexlab{}.
\newblock \showarticletitle{Multi-Objective Neural Architecture Search for In-Memory Computing}. In \bibinfo{booktitle}{\emph{2024 IEEE Computer Society Annual Symposium on VLSI (ISVLSI)}}. \bibinfo{pages}{343--348}.
\newblock
\href{https://doi.org/10.1109/ISVLSI61997.2024.00069}{doi:\nolinkurl{10.1109/ISVLSI61997.2024.00069}}


\bibitem[Ankit et~al\mbox{.}(2019)]%
        {PUMA}
\bibfield{author}{\bibinfo{person}{Aayush Ankit}, \bibinfo{person}{Izzat~El Hajj}, \bibinfo{person}{Sai~Rahul Chalamalasetti}, \bibinfo{person}{Geoffrey Ndu}, \bibinfo{person}{Martin Foltin}, \bibinfo{person}{R.~Stanley Williams}, \bibinfo{person}{Paolo Faraboschi}, \bibinfo{person}{Wen-mei~W Hwu}, \bibinfo{person}{John~Paul Strachan}, \bibinfo{person}{Kaushik Roy}, {and} \bibinfo{person}{Dejan~S. Milojicic}.} \bibinfo{year}{2019}\natexlab{}.
\newblock \showarticletitle{PUMA: A Programmable Ultra-Efficient Memristor-Based Accelerator for Machine Learning Inference}. In \bibinfo{booktitle}{\emph{Proceedings of the Twenty-Fourth International Conference on Architectural Support for Programming Languages and Operating Systems}} (Providence, RI, USA) \emph{(\bibinfo{series}{ASPLOS '19})}. \bibinfo{publisher}{Association for Computing Machinery}, \bibinfo{address}{New York, NY, USA}, \bibinfo{pages}{715–731}.
\newblock
\showISBNx{9781450362405}
\href{https://doi.org/10.1145/3297858.3304049}{doi:\nolinkurl{10.1145/3297858.3304049}}


\bibitem[Bender et~al\mbox{.}(2018)]%
        {oneshotzoph}
\bibfield{author}{\bibinfo{person}{Gabriel Bender}, \bibinfo{person}{Pieter-Jan Kindermans}, \bibinfo{person}{Barret Zoph}, \bibinfo{person}{Vijay Vasudevan}, {and} \bibinfo{person}{Quoc Le}.} \bibinfo{year}{2018}\natexlab{}.
\newblock \showarticletitle{Understanding and Simplifying One-Shot Architecture Search}. In \bibinfo{booktitle}{\emph{Proceedings of the 35th International Conference on Machine Learning}} \emph{(\bibinfo{series}{Proceedings of Machine Learning Research}, Vol.~\bibinfo{volume}{80})}, \bibfield{editor}{\bibinfo{person}{Jennifer Dy} {and} \bibinfo{person}{Andreas Krause}} (Eds.). \bibinfo{publisher}{PMLR}, \bibinfo{pages}{550--559}.
\newblock
\urldef\tempurl%
\url{https://proceedings.mlr.press/v80/bender18a.html}
\showURL{%
\tempurl}


\bibitem[Benmeziane et~al\mbox{.}(2023)]%
        {analogNAS}
\bibfield{author}{\bibinfo{person}{Hadjer Benmeziane}, \bibinfo{person}{Corey Lammie}, \bibinfo{person}{Irem Boybat}, \bibinfo{person}{Malte Rasch}, \bibinfo{person}{Manuel Le~Gallo}, \bibinfo{person}{Hsinyu Tsai}, \bibinfo{person}{Ramachandran Muralidhar}, \bibinfo{person}{Smail Niar}, \bibinfo{person}{Ouarnoughi Hamza}, \bibinfo{person}{Vijay Narayanan}, \bibinfo{person}{Abu Sebastian}, {and} \bibinfo{person}{Kaoutar El~Maghraoui}.} \bibinfo{year}{2023}\natexlab{}.
\newblock \showarticletitle{AnalogNAS: A Neural Network Design Framework for Accurate Inference with Analog In-Memory Computing}. In \bibinfo{booktitle}{\emph{2023 IEEE International Conference on Edge Computing and Communications (EDGE)}}. \bibinfo{pages}{233--244}.
\newblock
\href{https://doi.org/10.1109/EDGE60047.2023.00045}{doi:\nolinkurl{10.1109/EDGE60047.2023.00045}}


\bibitem[Cai et~al\mbox{.}(2019)]%
        {proxyless}
\bibfield{author}{\bibinfo{person}{Han Cai}, \bibinfo{person}{Ligeng Zhu}, {and} \bibinfo{person}{Song Han}.} \bibinfo{year}{2019}\natexlab{}.
\newblock \bibinfo{title}{ProxylessNAS: Direct Neural Architecture Search on Target Task and Hardware}.
\newblock
\showeprint[arxiv]{1812.00332}~[cs.LG]
\urldef\tempurl%
\url{https://arxiv.org/abs/1812.00332}
\showURL{%
\tempurl}


\bibitem[Chi et~al\mbox{.}(2016)]%
        {prime}
\bibfield{author}{\bibinfo{person}{Ping Chi}, \bibinfo{person}{Shuangchen Li}, \bibinfo{person}{Cong Xu}, \bibinfo{person}{Tao Zhang}, \bibinfo{person}{Jishen Zhao}, \bibinfo{person}{Yongpan Liu}, \bibinfo{person}{Yu Wang}, {and} \bibinfo{person}{Yuan Xie}.} \bibinfo{year}{2016}\natexlab{}.
\newblock \showarticletitle{PRIME: A Novel Processing-in-Memory Architecture for Neural Network Computation in ReRAM-Based Main Memory}. In \bibinfo{booktitle}{\emph{Proceedings of the 43rd International Symposium on Computer Architecture}} (Seoul, Republic of Korea) \emph{(\bibinfo{series}{ISCA ’16})}. \bibinfo{pages}{27–39}.
\newblock
\showISBNx{9781467389471}


\bibitem[Elbtity et~al\mbox{.}(2021)]%
        {iCASISVLSI22}
\bibfield{author}{\bibinfo{person}{Mohammed Elbtity}, \bibinfo{person}{Abhishek Singh}, \bibinfo{person}{Brendan Reidy}, \bibinfo{person}{Xiaochen Guo}, {and} \bibinfo{person}{Ramtin Zand}.} \bibinfo{year}{2021}\natexlab{}.
\newblock \showarticletitle{An In-Memory Analog Computing Co-Processor for Energy-Efficient CNN Inference on Mobile Devices}. In \bibinfo{booktitle}{\emph{2021 IEEE Computer Society Annual Symposium on VLSI (ISVLSI)}}. \bibinfo{pages}{188--193}.
\newblock
\href{https://doi.org/10.1109/ISVLSI51109.2021.00043}{doi:\nolinkurl{10.1109/ISVLSI51109.2021.00043}}


\bibitem[Elbtity et~al\mbox{.}(2023)]%
        {TPU-IMAC}
\bibfield{author}{\bibinfo{person}{Mohammed~E. Elbtity}, \bibinfo{person}{Brendan Reidy}, \bibinfo{person}{Md~Hasibul Amin}, {and} \bibinfo{person}{Ramtin Zand}.} \bibinfo{year}{2023}\natexlab{}.
\newblock \showarticletitle{Heterogeneous Integration of In-Memory Analog Computing Architectures with Tensor Processing Units}. In \bibinfo{booktitle}{\emph{Proceedings of the Great Lakes Symposium on VLSI 2023}} (Knoxville, TN, USA) \emph{(\bibinfo{series}{GLSVLSI '23})}. \bibinfo{publisher}{Association for Computing Machinery}, \bibinfo{address}{New York, NY, USA}, \bibinfo{pages}{607–612}.
\newblock
\showISBNx{9798400701252}
\href{https://doi.org/10.1145/3583781.3590256}{doi:\nolinkurl{10.1145/3583781.3590256}}


\bibitem[Guo et~al\mbox{.}(2020)]%
        {spos}
\bibfield{author}{\bibinfo{person}{Zichao Guo}, \bibinfo{person}{Xiangyu Zhang}, \bibinfo{person}{Haoyuan Mu}, \bibinfo{person}{Wen Heng}, \bibinfo{person}{Zechun Liu}, \bibinfo{person}{Yichen Wei}, {and} \bibinfo{person}{Jian Sun}.} \bibinfo{year}{2020}\natexlab{}.
\newblock \showarticletitle{Single Path One-Shot Neural Architecture Search with Uniform Sampling}. In \bibinfo{booktitle}{\emph{Computer Vision -- ECCV 2020}}, \bibfield{editor}{\bibinfo{person}{Andrea Vedaldi}, \bibinfo{person}{Horst Bischof}, \bibinfo{person}{Thomas Brox}, {and} \bibinfo{person}{Jan-Michael Frahm}} (Eds.). \bibinfo{publisher}{Springer International Publishing}, \bibinfo{address}{Cham}, \bibinfo{pages}{544--560}.
\newblock
\showISBNx{978-3-030-58517-4}


\bibitem[He et~al\mbox{.}(2016)]%
        {resnet}
\bibfield{author}{\bibinfo{person}{Kaiming He}, \bibinfo{person}{Xiangyu Zhang}, \bibinfo{person}{Shaoqing Ren}, {and} \bibinfo{person}{Jian Sun}.} \bibinfo{year}{2016}\natexlab{}.
\newblock \showarticletitle{Deep Residual Learning for Image Recognition}. In \bibinfo{booktitle}{\emph{2016 IEEE Conference on Computer Vision and Pattern Recognition (CVPR)}}. \bibinfo{pages}{770--778}.
\newblock
\href{https://doi.org/10.1109/CVPR.2016.90}{doi:\nolinkurl{10.1109/CVPR.2016.90}}


\bibitem[Hu et~al\mbox{.}(2016)]%
        {dpengine}
\bibfield{author}{\bibinfo{person}{Miao Hu}, \bibinfo{person}{John~Paul Strachan}, \bibinfo{person}{Zhiyong Li}, \bibinfo{person}{Emmanuelle~M. Grafals}, \bibinfo{person}{Noraica Davila}, \bibinfo{person}{Catherine Graves}, \bibinfo{person}{Sity Lam}, \bibinfo{person}{Ning Ge}, \bibinfo{person}{Jianhua~Joshua Yang}, {and} \bibinfo{person}{R.~Stanley Williams}.} \bibinfo{year}{2016}\natexlab{}.
\newblock \showarticletitle{Dot-product engine for neuromorphic computing: Programming 1T1M crossbar to accelerate matrix-vector multiplication}. In \bibinfo{booktitle}{\emph{2016 53nd ACM/EDAC/IEEE Design Automation Conference (DAC)}}. \bibinfo{pages}{1--6}.
\newblock
\href{https://doi.org/10.1145/2897937.2898010}{doi:\nolinkurl{10.1145/2897937.2898010}}


\bibitem[Jiang et~al\mbox{.}(2021)]%
        {nacim}
\bibfield{author}{\bibinfo{person}{Weiwen Jiang}, \bibinfo{person}{Qiuwen Lou}, \bibinfo{person}{Zheyu Yan}, \bibinfo{person}{Lei Yang}, \bibinfo{person}{Jingtong Hu}, \bibinfo{person}{Xiaobo~Sharon Hu}, {and} \bibinfo{person}{Yiyu Shi}.} \bibinfo{year}{2021}\natexlab{}.
\newblock \showarticletitle{Device-Circuit-Architecture Co-Exploration for Computing-in-Memory Neural Accelerators}.
\newblock \bibinfo{journal}{\emph{IEEE Trans. Comput.}} \bibinfo{volume}{70}, \bibinfo{number}{4} (\bibinfo{year}{2021}), \bibinfo{pages}{595--605}.
\newblock
\href{https://doi.org/10.1109/TC.2020.2991575}{doi:\nolinkurl{10.1109/TC.2020.2991575}}


\bibitem[Kim et~al\mbox{.}(2022)]%
        {IMCSurvey}
\bibfield{author}{\bibinfo{person}{Donghyuk Kim}, \bibinfo{person}{Chengshuo Yu}, \bibinfo{person}{Shanshan Xie}, \bibinfo{person}{Yuzong Chen}, \bibinfo{person}{Joo-Young Kim}, \bibinfo{person}{Bongjin Kim}, \bibinfo{person}{Jaydeep~P. Kulkarni}, {and} \bibinfo{person}{Tony Tae-Hyoung Kim}.} \bibinfo{year}{2022}\natexlab{}.
\newblock \showarticletitle{An Overview of Processing-in-Memory Circuits for Artificial Intelligence and Machine Learning}.
\newblock \bibinfo{journal}{\emph{IEEE Journal on Emerging and Selected Topics in Circuits and Systems}} \bibinfo{volume}{12}, \bibinfo{number}{2} (\bibinfo{year}{2022}), \bibinfo{pages}{338--353}.
\newblock
\href{https://doi.org/10.1109/JETCAS.2022.3160455}{doi:\nolinkurl{10.1109/JETCAS.2022.3160455}}


\bibitem[Kim et~al\mbox{.}(2019)]%
        {variation}
\bibfield{author}{\bibinfo{person}{Sungho Kim}, \bibinfo{person}{Hee-Dong Kim}, {and} \bibinfo{person}{Sung-Jin Choi}.} \bibinfo{year}{2019}\natexlab{}.
\newblock \showarticletitle{impact of Synaptic Device Variations on Classification Accuracy in a Binarized neural network}.
\newblock \bibinfo{journal}{\emph{Scientific reports}} \bibinfo{volume}{9}, \bibinfo{number}{1} (\bibinfo{year}{2019}), \bibinfo{pages}{1--7}.
\newblock


\bibitem[Kingma and Ba(2017)]%
        {adam}
\bibfield{author}{\bibinfo{person}{Diederik~P. Kingma} {and} \bibinfo{person}{Jimmy Ba}.} \bibinfo{year}{2017}\natexlab{}.
\newblock \bibinfo{title}{Adam: A Method for Stochastic Optimization}.
\newblock
\showeprint[arxiv]{1412.6980}~[cs.LG]
\urldef\tempurl%
\url{https://arxiv.org/abs/1412.6980}
\showURL{%
\tempurl}


\bibitem[Krizhevsky and Hinton(2009)]%
        {cifar}
\bibfield{author}{\bibinfo{person}{Alex Krizhevsky} {and} \bibinfo{person}{Geoffrey Hinton}.} \bibinfo{year}{2009}\natexlab{}.
\newblock \bibinfo{booktitle}{\emph{Learning multiple layers of features from tiny images}}.
\newblock \bibinfo{type}{{T}echnical {R}eport}~0. \bibinfo{institution}{University of Toronto}, \bibinfo{address}{Toronto, Ontario}.
\newblock
\urldef\tempurl%
\url{https://www.cs.toronto.edu/~kriz/learning-features-2009-TR.pdf}
\showURL{%
\tempurl}


\bibitem[Krizhevsky et~al\mbox{.}(2012)]%
        {alexnet}
\bibfield{author}{\bibinfo{person}{Alex Krizhevsky}, \bibinfo{person}{Ilya Sutskever}, {and} \bibinfo{person}{Geoffrey~E Hinton}.} \bibinfo{year}{2012}\natexlab{}.
\newblock \showarticletitle{ImageNet Classification with Deep Convolutional Neural Networks}. In \bibinfo{booktitle}{\emph{Advances in Neural Information Processing Systems}}, \bibfield{editor}{\bibinfo{person}{F.~Pereira}, \bibinfo{person}{C.J. Burges}, \bibinfo{person}{L.~Bottou}, {and} \bibinfo{person}{K.Q. Weinberger}} (Eds.), Vol.~\bibinfo{volume}{25}. \bibinfo{publisher}{Curran Associates, Inc.}
\newblock
\urldef\tempurl%
\url{https://proceedings.neurips.cc/paper_files/paper/2012/file/c399862d3b9d6b76c8436e924a68c45b-Paper.pdf}
\showURL{%
\tempurl}


\bibitem[Le~Gallo et~al\mbox{.}(2018)]%
        {mixedIMC}
\bibfield{author}{\bibinfo{person}{Manuel Le~Gallo}, \bibinfo{person}{Abu Sebastian}, \bibinfo{person}{Roland Mathis}, \bibinfo{person}{Matteo Manica}, \bibinfo{person}{Heiner Giefers}, \bibinfo{person}{Tomas Tuma}, \bibinfo{person}{Costas Bekas}, \bibinfo{person}{Alessandro Curioni}, {and} \bibinfo{person}{Evangelos Eleftheriou}.} \bibinfo{year}{2018}\natexlab{}.
\newblock \showarticletitle{Mixed-precision in-memory computing}.
\newblock \bibinfo{journal}{\emph{Nature Electronics}} \bibinfo{volume}{1}, \bibinfo{number}{4} (\bibinfo{year}{2018}), \bibinfo{pages}{246--253}.
\newblock


\bibitem[Negi et~al\mbox{.}(2022)]%
        {nax}
\bibfield{author}{\bibinfo{person}{Shubham Negi}, \bibinfo{person}{Indranil Chakraborty}, \bibinfo{person}{Aayush Ankit}, {and} \bibinfo{person}{Kaushik Roy}.} \bibinfo{year}{2022}\natexlab{}.
\newblock \showarticletitle{NAX: neural architecture and memristive xbar based accelerator co-design}. In \bibinfo{booktitle}{\emph{Proceedings of the 59th ACM/IEEE Design Automation Conference}} (San Francisco, California) \emph{(\bibinfo{series}{DAC '22})}. \bibinfo{publisher}{Association for Computing Machinery}, \bibinfo{address}{New York, NY, USA}, \bibinfo{pages}{451–456}.
\newblock
\showISBNx{9781450391429}
\href{https://doi.org/10.1145/3489517.3530476}{doi:\nolinkurl{10.1145/3489517.3530476}}


\bibitem[Pham et~al\mbox{.}(2018)]%
        {wsnas-basic}
\bibfield{author}{\bibinfo{person}{Hieu Pham}, \bibinfo{person}{Melody Guan}, \bibinfo{person}{Barret Zoph}, \bibinfo{person}{Quoc Le}, {and} \bibinfo{person}{Jeff Dean}.} \bibinfo{year}{2018}\natexlab{}.
\newblock \showarticletitle{Efficient Neural Architecture Search via Parameters Sharing}. In \bibinfo{booktitle}{\emph{Proceedings of the 35th International Conference on Machine Learning}} \emph{(\bibinfo{series}{Proceedings of Machine Learning Research}, Vol.~\bibinfo{volume}{80})}, \bibfield{editor}{\bibinfo{person}{Jennifer Dy} {and} \bibinfo{person}{Andreas Krause}} (Eds.). \bibinfo{publisher}{PMLR}, \bibinfo{pages}{4095--4104}.
\newblock
\urldef\tempurl%
\url{https://proceedings.mlr.press/v80/pham18a.html}
\showURL{%
\tempurl}


\bibitem[Ruder(2017)]%
        {sgd}
\bibfield{author}{\bibinfo{person}{Sebastian Ruder}.} \bibinfo{year}{2017}\natexlab{}.
\newblock \bibinfo{title}{An overview of gradient descent optimization algorithms}.
\newblock
\showeprint[arxiv]{1609.04747}~[cs.LG]
\urldef\tempurl%
\url{https://arxiv.org/abs/1609.04747}
\showURL{%
\tempurl}


\bibitem[Shafiee et~al\mbox{.}(2016)]%
        {ISAAC}
\bibfield{author}{\bibinfo{person}{Ali Shafiee}, \bibinfo{person}{Anirban Nag}, \bibinfo{person}{Naveen Muralimanohar}, \bibinfo{person}{Rajeev Balasubramonian}, \bibinfo{person}{John~Paul Strachan}, \bibinfo{person}{Miao Hu}, \bibinfo{person}{R.~Stanley Williams}, {and} \bibinfo{person}{Vivek Srikumar}.} \bibinfo{year}{2016}\natexlab{}.
\newblock \showarticletitle{ISAAC: A Convolutional Neural Network Accelerator with in-Situ Analog Arithmetic in Crossbars} \emph{(\bibinfo{series}{ISCA '16})}. \bibinfo{publisher}{IEEE Press}, \bibinfo{pages}{14–26}.
\newblock
\showISBNx{9781467389471}
\href{https://doi.org/10.1109/ISCA.2016.12}{doi:\nolinkurl{10.1109/ISCA.2016.12}}


\bibitem[Simonyan and Zisserman(2015)]%
        {vgg16}
\bibfield{author}{\bibinfo{person}{Karen Simonyan} {and} \bibinfo{person}{Andrew Zisserman}.} \bibinfo{year}{2015}\natexlab{}.
\newblock \showarticletitle{Very Deep Convolutional Networks for Large-Scale Image Recognition}. In \bibinfo{booktitle}{\emph{International Conference on Learning Representations}}.
\newblock


\bibitem[Stamoulis et~al\mbox{.}(2020)]%
        {spos-hw}
\bibfield{author}{\bibinfo{person}{Dimitrios Stamoulis}, \bibinfo{person}{Ruizhou Ding}, \bibinfo{person}{Di Wang}, \bibinfo{person}{Dimitrios Lymberopoulos}, \bibinfo{person}{Bodhi Priyantha}, \bibinfo{person}{Jie Liu}, {and} \bibinfo{person}{Diana Marculescu}.} \bibinfo{year}{2020}\natexlab{}.
\newblock \showarticletitle{Single-Path NAS: Designing Hardware-Efficient ConvNets in Less Than 4 Hours}. In \bibinfo{booktitle}{\emph{Machine Learning and Knowledge Discovery in Databases}}, \bibfield{editor}{\bibinfo{person}{Ulf Brefeld}, \bibinfo{person}{Elisa Fromont}, \bibinfo{person}{Andreas Hotho}, \bibinfo{person}{Arno Knobbe}, \bibinfo{person}{Marloes Maathuis}, {and} \bibinfo{person}{C{\'e}line Robardet}} (Eds.). \bibinfo{publisher}{Springer International Publishing}, \bibinfo{address}{Cham}, \bibinfo{pages}{481--497}.
\newblock
\showISBNx{978-3-030-46147-8}


\bibitem[Sun et~al\mbox{.}(2020)]%
        {PIM-quant}
\bibfield{author}{\bibinfo{person}{Hanbo Sun}, \bibinfo{person}{Zhenhua Zhu}, \bibinfo{person}{Yi Cai}, \bibinfo{person}{Xiaoming Chen}, \bibinfo{person}{Yu Wang}, {and} \bibinfo{person}{Huazhong Yang}.} \bibinfo{year}{2020}\natexlab{}.
\newblock \showarticletitle{An Energy-Efficient Quantized and Regularized Training Framework For Processing-In-Memory Accelerators}. In \bibinfo{booktitle}{\emph{2020 25th Asia and South Pacific Design Automation Conference (ASP-DAC)}}. \bibinfo{pages}{325--330}.
\newblock
\href{https://doi.org/10.1109/ASP-DAC47756.2020.9045192}{doi:\nolinkurl{10.1109/ASP-DAC47756.2020.9045192}}


\bibitem[Sun et~al\mbox{.}(2023)]%
        {gibbon}
\bibfield{author}{\bibinfo{person}{Hanbo Sun}, \bibinfo{person}{Zhenhua Zhu}, \bibinfo{person}{Chenyu Wang}, \bibinfo{person}{Xuefei Ning}, \bibinfo{person}{Guohao Dai}, \bibinfo{person}{Huazhong Yang}, {and} \bibinfo{person}{Yu Wang}.} \bibinfo{year}{2023}\natexlab{}.
\newblock \showarticletitle{Gibbon: An Efficient Co-Exploration Framework of NN Model and Processing-In-Memory Architecture}.
\newblock \bibinfo{journal}{\emph{IEEE Transactions on Computer-Aided Design of Integrated Circuits and Systems}} \bibinfo{volume}{42}, \bibinfo{number}{11} (\bibinfo{year}{2023}), \bibinfo{pages}{4075--4089}.
\newblock
\href{https://doi.org/10.1109/TCAD.2023.3262201}{doi:\nolinkurl{10.1109/TCAD.2023.3262201}}


\bibitem[Wu et~al\mbox{.}(2019)]%
        {fbnet}
\bibfield{author}{\bibinfo{person}{Bichen Wu}, \bibinfo{person}{Xiaoliang Dai}, \bibinfo{person}{Peizhao Zhang}, \bibinfo{person}{Yanghan Wang}, \bibinfo{person}{Fei Sun}, \bibinfo{person}{Yiming Wu}, \bibinfo{person}{Yuandong Tian}, \bibinfo{person}{Peter Vajda}, \bibinfo{person}{Yangqing Jia}, {and} \bibinfo{person}{Kurt Keutzer}.} \bibinfo{year}{2019}\natexlab{}.
\newblock \showarticletitle{FBNet: Hardware-Aware Efficient ConvNet Design via Differentiable Neural Architecture Search}. In \bibinfo{booktitle}{\emph{2019 IEEE/CVF Conference on Computer Vision and Pattern Recognition (CVPR)}}. \bibinfo{pages}{10726--10734}.
\newblock
\href{https://doi.org/10.1109/CVPR.2019.01099}{doi:\nolinkurl{10.1109/CVPR.2019.01099}}


\bibitem[Yan et~al\mbox{.}(2021)]%
        {uae}
\bibfield{author}{\bibinfo{person}{Zheyu Yan}, \bibinfo{person}{Da-Cheng Juan}, \bibinfo{person}{Xiaobo~Sharon Hu}, {and} \bibinfo{person}{Yiyu Shi}.} \bibinfo{year}{2021}\natexlab{}.
\newblock \showarticletitle{Uncertainty Modeling of Emerging Device based Computing-in-Memory Neural Accelerators with Application to Neural Architecture Search}. In \bibinfo{booktitle}{\emph{2021 26th Asia and South Pacific Design Automation Conference (ASP-DAC)}}. \bibinfo{pages}{859--864}.
\newblock


\bibitem[Yuan et~al\mbox{.}(2021)]%
        {nas4rram}
\bibfield{author}{\bibinfo{person}{Zhihang Yuan}, \bibinfo{person}{Jingze Liu}, \bibinfo{person}{Xingchen Li}, \bibinfo{person}{Longhao Yan}, \bibinfo{person}{Haoxiang Chen}, \bibinfo{person}{Bingzhe Wu}, \bibinfo{person}{Yuchao Yang}, {and} \bibinfo{person}{Guangyu Sun}.} \bibinfo{year}{2021}\natexlab{}.
\newblock \showarticletitle{NAS4RRAM: neural network architecture search for inference on RRAM-based accelerators}.
\newblock \bibinfo{journal}{\emph{Science China Information Sciences}} \bibinfo{volume}{64}, \bibinfo{number}{6} (\bibinfo{date}{10 May} \bibinfo{year}{2021}), \bibinfo{pages}{160407}.
\newblock
\showISSN{1869-1919}
\href{https://doi.org/10.1007/s11432-020-3245-7}{doi:\nolinkurl{10.1007/s11432-020-3245-7}}


\bibitem[Zand et~al\mbox{.}(2018)]%
        {zand2018fundamentals}
\bibfield{author}{\bibinfo{person}{Ramtin Zand}, \bibinfo{person}{Arman Roohi}, {and} \bibinfo{person}{Ronald~F DeMara}.} \bibinfo{year}{2018}\natexlab{}.
\newblock \showarticletitle{Fundamentals, modeling, and application of magnetic tunnel junctions}.
\newblock In \bibinfo{booktitle}{\emph{Nanoscale Devices}}. \bibinfo{publisher}{CRC Press}, \bibinfo{pages}{337--368}.
\newblock


\bibitem[Zhang et~al\mbox{.}(2021)]%
        {ssnas}
\bibfield{author}{\bibinfo{person}{Xinbang Zhang}, \bibinfo{person}{Zehao Huang}, \bibinfo{person}{Naiyan Wang}, \bibinfo{person}{Shiming Xiang}, {and} \bibinfo{person}{Chunhong Pan}.} \bibinfo{year}{2021}\natexlab{}.
\newblock \showarticletitle{{ You Only Search Once: Single Shot Neural Architecture Search via Direct Sparse Optimization }}.
\newblock \bibinfo{journal}{\emph{IEEE Transactions on Pattern Analysis \& Machine Intelligence}} \bibinfo{volume}{43}, \bibinfo{number}{09} (\bibinfo{date}{Sept.} \bibinfo{year}{2021}), \bibinfo{pages}{2891--2904}.
\newblock
\showISSN{1939-3539}
\href{https://doi.org/10.1109/TPAMI.2020.3020300}{doi:\nolinkurl{10.1109/TPAMI.2020.3020300}}


\bibitem[Zhu et~al\mbox{.}(2023)]%
        {MNSIM2}
\bibfield{author}{\bibinfo{person}{Zhenhua Zhu}, \bibinfo{person}{Hanbo Sun}, \bibinfo{person}{Tongxin Xie}, \bibinfo{person}{Yu Zhu}, \bibinfo{person}{Guohao Dai}, \bibinfo{person}{Lixue Xia}, \bibinfo{person}{Dimin Niu}, \bibinfo{person}{Xiaoming Chen}, \bibinfo{person}{Xiaobo~Sharon Hu}, \bibinfo{person}{Yu Cao}, \bibinfo{person}{Yuan Xie}, \bibinfo{person}{Huazhong Yang}, {and} \bibinfo{person}{Yu Wang}.} \bibinfo{year}{2023}\natexlab{}.
\newblock \showarticletitle{MNSIM 2.0: A Behavior-Level Modeling Tool for Processing-In-Memory Architectures}.
\newblock \bibinfo{journal}{\emph{IEEE Transactions on Computer-Aided Design of Integrated Circuits and Systems}} \bibinfo{volume}{42}, \bibinfo{number}{11} (\bibinfo{year}{2023}), \bibinfo{pages}{4112--4125}.
\newblock
\href{https://doi.org/10.1109/TCAD.2023.3251696}{doi:\nolinkurl{10.1109/TCAD.2023.3251696}}


\end{thebibliography}

\end{document}